\documentclass[twocolumn,showpacs,showkeys,preprintnumbers,amsmath,amssymb]{revtex4}

\usepackage{graphicx}
\usepackage{dcolumn}
\usepackage{bm}

\newcommand{\be}{\begin{equation}}
\newcommand{\ee}{\end{equation}}
\newcommand{\ba}{\begin{eqnarray}}
\newcommand{\ea}{\end{eqnarray}}

\def\gs{\mathrel{\raise1.16pt\hbox{$>$}\kern-7.0pt %
\lower3.06pt\hbox{{$\scriptstyle \sim$}}}}         %
\def\ls{\mathrel{\raise1.16pt\hbox{$<$}\kern-7.0pt %
\lower3.06pt\hbox{{$\scriptstyle \sim$}}}}         %

\begin{document}

\preprint{APS/000-000}

\title{Finding Evidence for Massive Neutrinos using 3D Weak Lensing}

\author{T. D. Kitching}
\email{tdk@astro.ox.ac.uk}
 \affiliation{University of Oxford, Denys Wilkinson Building, Department of Physics, 
Wilkinson Building, Keble Road, Oxford OX1 3RH, UK}

\author{A. F. Heavens}%
\affiliation{SUPA\footnote{Scottish Universities Physics Alliance} Institute for Astronomy, University of Edinburgh, Royal
Observatory, Blackford Hill, Edinburgh, EH9 3HJ,UK
}%

\author{L. Verde}
\affiliation{ICREA \& Institute of Space Sciences (IEEC-CSIC),Fac. Ciencies, 
Campus UAB Torre C5 parell 2 Bellaterra, Spain
}%

\author{P. Serra}
\affiliation{Department of Physics and Astronomy,
University of California, Irvine, CA 92617, USA
}%

\author{A. Melchiorri}
\affiliation{Dipartimento di Fisica e Sezione INFN, Universita' degli Studi di Roma \\
``La Sapienza'', Ple Aldo Moro 5,00185, Rome, Italy
}%

\begin{abstract}

In this paper we investigate the potential of 3D cosmic shear 
to constrain massive neutrino parameters. We find that if the total mass is substantial
(near the upper limits from LSS, but setting aside the Ly alpha limit
for now), then 3D cosmic shear + \emph{Planck} is very sensitive to
neutrino mass and one may expect
that a next generation photometric redshift survey could constrain the number of neutrinos 
$N_{\nu}$ and the sum of their masses $m_{\nu}=\sum_i m_i$ to 
an accuracy of $\Delta N_{\nu}\sim 0.08$ and $\Delta m_{\nu}\sim 0.03$ eV respectively.
If in fact the masses are close to zero, then the
errors weaken to $\Delta N_{\nu}\sim 0.10$ and $\Delta m_{\nu}\sim 0.07$ eV. In either 
case there is a factor $4$ improvement over \emph{Planck} alone.
We use a Bayesian evidence method to predict \emph{joint} expected 
evidence for $N_{\nu}$ and $m_{\nu}$. We find that 3D cosmic shear combined with a 
\emph{Planck} prior could provide `substantial' evidence for massive neutrinos 
and be able to 
distinguish `decisively' between many
competing massive neutrino models. This technique should `decisively' 
distinguish between models in which there are no massive neutrinos and models in which 
there are massive neutrinos with
$|N_\nu-3|\gs 0.35$ and $m_{\nu}\gs 0.25$ eV.  
We introduce the notion of marginalised and conditional 
evidence when considering evidence for individual parameter values within a multi-parameter model. 

\end{abstract}



\maketitle

\section{Introduction}
\label{Introduction}
In this paper we will investigate the potential of 3D weak lensing to constrain
the properties of, and provide evidence for, massive neutrinos. The conclusion that 
neutrinos have mass, and the 
resolution of the actual masses, would have a profound impact on our understanding of 
particle physics and 
cosmology.

As a photon travels from a distant galaxy the path it takes is diverted, by the presence of large-scale 
structure along the line of sight resulting in the image of any galaxy being slightly distorted. This 
weak lensing effect depends on both the details of the matter power spectrum and growth of structure as well 
as the geometry of the observer-lens-source configuration. 
3D weak lensing combines this weak lensing information 
with any redshift information available which then allows for evolving effects, for example dark energy, to be 
investigated. Weak lensing (see \cite{art:Munshi} for a recent review) 
has been used to constrain cosmological 
parameters including dark energy parameters \cite{art:Kitching07}, 
measure the growth of structure \cite{art:Massey,art:Bacon} and map the dark matter distribution as a 
function of redshift \cite{art:Taylor}. 
It has been shown that 3D weak lensing has the potential to constrain dark energy parameters to an unprecedented degree of accuracy using upcoming and 
future surveys e.g. Pan-STARRS \cite{art:Kaiser} 
and \emph{DUNE} 
\cite{art:Refregier}. In addition to dark energy parameters, galaxy redshift surveys \cite{art:Hannestad07, art:Abdalla, art:Takada} and 
weak lensing tomography (in which galaxies are binned in redshift)~\cite{art:Hannestad06, art:Abazajian, art:Cooray} have the potential to constrain the total neutrino mass. 
In this paper we consider a novel technique, 3D cosmic shear \cite{art:Castro, art:Heavens03, art:Heavens06, art:Kitching07} 
in which galaxies are not binned in redshift and the full 3D shear field is used, thus maximising the information extracted. We the report 3D cosmic shear 
performance in constraining neutrino properties and also present 
Bayesian evidence calculations that will show whether future weak lensing surveys will find convincing evidence for 
massive neutrinos. 

In the standard model of particle physics neutrinos must have zero mass by definition; if neutrinos have mass this 
may be a signal of non-standard neutrino interactions or Higgs mechanisms, evidence that the standard model is 
not renormalisable, or extra (stringy) dimensions (we refer to neutrino mass and 
oscillation reviews \cite{art:King, art:Lesgourgues, art:PDG} and references therein). 
In cosmology neutrinos play a role in structure formation by 
damping structure on small scales. They can be categorised as a hot 
component of dark matter, though recent results 
(for example WMAP \cite{art:Spergel}) rule out massive neutrinos 
as the dominant dark matter component. 
Beyond the affect of neutrinos on large scale structure they are of further interest in cosmology since 
by directly observing neutrinos they could  
provide a window on the early Universe beyond the surface of 
last scattering to the epoch of electroweak unification. Theoretically 
massive (majorana; $\nu_{\alpha}=\bar\nu_{\alpha}$) neutrinos may provide some explanation for the 
baryon asymmetry \cite{art:Fukugita}. And indeed astronomical considerations first 
alluded to neutrino mass since the number of neutrinos detected from the Sun was much less than the expected 
number from the Sun's luminosity. 

There is currently a substantial and growing amount of evidence that neutrinos have mass (for recent reviews 
see \cite{art:King, art:Lesgourgues, art:PDG}). This conclusion has 
been reached using results from large particle physics experiments which have found that the oscillation of 
neutrinos from one flavour ($e$, $\mu$ or $\tau$) to another is needed to explain the observed data. 
SuperKamiokande \cite{art:Fukuda} found that only $\sim 1/3$ the flux of muon neutrinos $\nu_{\mu}$ 
from cosmic ray collisions in 
the atmosphere were observed along the line-of-sight through the Earth implying an 
oscillation of $\nu_{\mu}$ to some other flavour with a scale length comparable to the radius of the Earth. 
SNO \cite{art:Ahmed} has observed both the total
flux of neutrinos from the Sun as well as the flux of electron neutrinos $\nu_{e}$ and found that the `solar 
neutrino problem' is resolved by postulating that $\nu_{e}$ oscillate to other flavours in the high density 
environment of the Sun's core by gaining a small effective mass. Further evidence for neutrino oscillations 
comes from nuclear reactors (KamLAND, \cite{art:Eguchi}) and neutrino beam experiments 
(K2K, \cite{art:Ahn}). 

The observed oscillation of neutrinos is linked to the implication that neutrinos have mass via the 
lepton mixing matrix $U^2$ whose elements describe the probability for one neutrino flavour to 
oscillate to another
\be
|\nu_\alpha\rangle=U_{\alpha i}|\nu_i\rangle;
\ee
this relates the observed neutrino flavours $|\nu_\alpha\rangle$ to a hierarchy of neutrino mass states $|\nu_i\rangle$. In general the elements of $U$ 
can be complex. The standard model has $3$ neutrino flavours $\nu_\alpha=\nu_{e}$, 
$\nu_{\mu}$, $\nu_{\tau}$. 
The lepton mixing matrix allows for more than $3$ neutrino flavours; if there 
are more than $3$ then the extra neutrinos are called ``sterile'' since they would not 
couple to any standard model electroweak interaction (i.e. would not be associated with electrons, muons or tau). 

Assuming $3$ neutrino flavours 
$U$ can be parameterised as a product of three Euler rotations $U=R_{23}R_{13}R_{12}$ where each 
rotation describes how one neutrino mass state is coupled to another. The elements of the rotation matrices depend 
on the \emph{mass difference} $\delta m_{ij}=m_i-m_j$ and an angle $\theta_{ij}$ on which the probability 
$P(\nu_i\rightarrow \nu_j)$ depends. Thus neutrino oscillation experiments can only measure the relative masses of
neutrinos not the absolute mass scale. 

Currently world neutrino data are consistent with a three-flavour mixing
framework (see \cite{fogli04} and references therein), parameterised in
terms of three neutrino masses $(m_1,m_2,m_3)$ and of three mixing
angles $(\theta_{12},\theta_{23},\theta_{13})$, plus a possible CP violating phase $\delta$.
Current constraints are $\delta m_{23} \sim 0.05$ eV and 
$\delta m_{12} \sim 0.007$ eV. There are currently no strong constraints on 
$\delta m_{13}$ though upcoming experiments, 
for example T2K \cite{art:Itow} should measure $\Delta\theta_{13}\sim 0.05$. Thus current constraints allow for two 
possible orders of the massive neutrino hierarchy: $m_1 < m_2 < m_3$ or the inverted hierarchy 
$m_3 < m_1 < m_2$. There are planned particle physics experiments that will measure the absolute mass scale via 
the beta decay of Tritium to constrain $m_{\nu e}$ \cite{Mbet,Main,Troi}, for example KATRIN \cite{art:Osipowski, art:Kristiansen} should reach an accuracy 
of $\Delta m_{\nu e}\sim 0.35$ eV. There are arguments (see 
\cite{art:Lesgourgues}) that the requirement on the accuracy of the absolute mass scale needed to break 
the hierarchy or inverted-hierarchy degeneracy is $\Delta m_{\nu}\ls 0.1$ eV. As discussed below this can be achieved by cosmological observations. 

Cosmologically massive neutrinos play a role in structure formation since free-streaming neutrinos can suppress 
growth on small scales. Neutrinos streaming from an over-density will reduce the amount of matter that can 
gravitationally accumulate by providing an extra effective pressure. In the nomenclature of cosmological 
parameter estimation massive neutrinos modify the matter power spectrum's growth rate by providing a suppression at 
small scales. It can thus leave key signatures in large scale structure data 
(see, eg.,\cite{art:Eisenstein}) and, to a lesser extent, in CMB data 
(see, e.g.,\cite{Ma95}). Very recently, it has also been shown that accurate
Lyman-$\alpha$ (Ly$\alpha$) forest data \cite{Mc04}, taken at face
value, can improve the current CMB+LSS constraints on $m_{\nu}$ by a
factor of $\sim 3$, with important consequences on absolute neutrino
mass scenarios \cite{art:Seljaks}. 
Current cosmological constraints from WMAP CMB combined with SDSS BAO and including 
Lyman-$\alpha$ constraints provide an current upper limit 
on the total mass 
of $m_{\nu}\ls 0.42$--$0.79$ eV depending on the assumptions made, with a median value of 
$m_{\nu}\ls 0.67$ eV \cite{art:Spergel, art:Seljak05, 
art:Peiris, art:Loveday, art:Fogli:2006yq}.  

There has been substantial work in numerically 
estimating the growth rate including the presence of 
massive neutrinos (for example \cite{art:Eisenstein, art:Kiakotou}). A degeneracy 
between neutrino mass 
and dark energy parameters arises because they both effectively suppress the matter power spectrum growth rate as highlighted by \cite{art:Kiakotou}. 
Optimistically, methods which can constrain dark energy parameters well 
should also be able to constrain the neutrino mass parameters, and by combining constraints from multiple methods 
(e.g. CMB, weak lensing) parameter degeneracies should be lifted. 

It is currently believed that the hierarchical mass scale of neutrinos implies that the total mass of neutrinos will
be approximately $m_{\nu}\sim 0.04$ -- $0.1$ 
therefore probes which are sensitive to this range of values are 
required to effectively constrain the neutrino mass. 
The remainder of this paper will highlight the possibility of using 
3D cosmic shear to constrain massive neutrino properties. We will introduce the 
methodology and assumptions made in Section \ref{Methodology} and present 
results in Sections \ref{Results} and \ref{Bayesian Evidence Results}, 
in Section \ref{Conclusion} we will discuss our conclusions. 

\section{Methodology}
\label{Methodology}
The central quantity in cosmological neutrino mass constraints 
is the fraction of the total matter density that is attributed to massive neutrinos 
$f_{\nu}\equiv\Omega_{\nu}/\Omega_m$ which we take from \cite{art:Takada} (Eqn. 1) 
\be
f_{\nu}=0.05\left(\frac{m_{\nu}}{0.658{\rm eV}}\right)\left(\frac{0.14}{\Omega_m h^2}\right)
\ee
where  $m_{\nu}=\sum_{i=0}^{N_{\nu}}m_i$ is the total neutrino mass, a sum over all 
neutrino species each with a mass $m_i$. The effect of massive neutrinos on the matter power spectrum is commonly 
expressed using $\Delta P(k)/P(k)=[P(k; f_{\nu})-P(k; f_{\nu}=0)]/P(k; f_{\nu}=0)$ which decreases and is negative 
as the wave-number $k$ increases and power is suppressed due to the free-streaming of the neutrinos. 
We use the Eisenstein \& Hu (1998) \cite{art:Eisenstein} fitting formula for the linear power spectrum 
which depends on both the number of massive neutrino species $N_{\nu}$ 
and the total neutrino mass and use the modification of the linear growth 
factor $f\equiv d\ln\delta/d\ln a$ suggested by \cite{art:Kiakotou} 
which adds a further dependence on $f_{\nu}$ and in addition
a dependence on the dark energy equation of state $w\equiv p/\rho$. 
We use the common parameterisation 
of the dark energy equation of state $w(a)=w_0+w_a(1-a)$ \cite{art:Chevallier}. 
One limitation of the Eisenstein \& Hu (1998) \cite{art:Eisenstein} parameterisation is that it assumes that each massive neutrino 
has the same mass i.e. $m_{\nu}=N_{\nu}m_i$ where $m_i$ is the same for all neutrino mass eigenstates. 
Further to this approximation we will also treat the number of massive neutrinos as a continuous 
parameter that is fitted by data, as opposed to an integer number, this can be justified since 
any light particle that does not 
couple to electrons, photons or ions will contribute to the effective number. However these assumptions 
imply that the constraints and 
predicted evidence presented are meant to be indicative of the ability of 3D cosmic shear to constrain 
neutrino mass and not as entirely representative of the situation as it is currently understood -- in which 
there are an integer number of mass eigenstates each with a different mass. 

We 
use a $k$ range of $k=0.001$ -- $1$ Mpc$^{-1}$ for the weak lensing Fisher matrix calculations 
and use the Smith et al. (2003) \cite{art:Smith} 
non-linear correction to the linear power spectrum. 
Note that this is in the quasi-linear
r\'egime and using wavenumbers that are at the limit of the reliability of the 
linear power spectrum fitting formula for massive neutrinos, for a recent review of the effect 
of massive neutrinos on the non-linear power spectrum see \cite{art:Saito}.

\subsection{3D Cosmic Shear}
The cosmological probes that we will consider in this paper are 3D cosmic shear and CMB. We 
use a CMB \emph{Planck} Fisher matrix 
which is calculated using {\sc CMBfast} (version 
4.5.1, \cite{art:Seljak96}) using the method outlined in \cite{art:Taylor} and will 
effectively be used as a prior in the results presented for 3D cosmic shear. For a general 
discussion of using \emph{Planck} to constrain neutrino mass parameters see \cite{art:Free}.

The weak lensing method we use, 3D cosmic shear \cite{art:Heavens03, art:Castro, art:Heavens06, art:Kitching07}, uses the weak lensing shear and 
redshift information of 
every galaxy. The 3D shear field is expanded in spherical harmonics and  
the covariance of the transform 
coefficients can be used to constrain cosmological parameters. 
The transform coefficients for a given set of azimuthal $\ell$ and radial $k$ 
[$h$Mpc$^{-1}$] wave numbers are given by summing over all galaxies $g$; 
\be
\label{3dshear}
\hat\gamma(k,\ell)=
\sum_g \gamma^g kj_{\ell}(k r^g){\rm e}^{-i\ell.\theta^g},
\ee
following the conventions of \cite{art:Castro}, and we assume a flat sky approximation. $\theta^g$ is the 
angular position of a galaxy on the sky, $r^g$ is the comoving distance to the galaxy and 
$j_{\ell}$ are spherical Bessel functions. $\gamma^g$ is the measured shear of the galaxy which 
parameterises the amount of distortion that the galaxy image has obtained due to intervening large scale structure. 

This is a novel approach over other 3D weak lensing analyses since galaxies 
are not binned in redshift which may cause problems at the bin boundaries and mean information loss in averaging over the bins. 
The binning approach, weak lensing tomography, creates a 2D map of the galaxies distortions at each 
redshift and takes the cross-correlation between each map to gain some extra 3D information. Conversely, 3D cosmic 
shear, presented here, uses the entire 3D shear field thus maximising the potential for 
information to be extracted from the galaxy image distortions. 

Since the mean of the coefficients in Eqn. (\ref{3dshear}) is zero the covariance is varied until it matches that of the data 
\cite{art:Kitching07}, i.e. the covariance is used as the `signal'. 
The 3D cosmic shear covariance 
depends on the the lensing geometry and the matter power spectrum, so the total parameter set
that can be constrained is:
$\Omega_m$, $\Omega_{de}$, $\Omega_b$, $h$, $\sigma_8$, $w_0$, $w_a$, $n_s$, 
the running of the spectral index $\alpha_n$, $m_{\nu}$ and $N_{\nu}$. We also include the 
tensor to scalar ratio $r=T/S$ and the optical depth to last scattering $\tau$ for the CMB Fisher 
matrix calculation. We do not assume spatial flatness and 
all results on individual parameters are fully marginalised over all other cosmological parameters. 
We use an $\ell_{\rm max}=5000$ for the 3D cosmic shear analysis and a $k_{\rm max}=1.0$ Mpc$^{-1}$ 
and use the same assumptions presented in \cite{art:Heavens06}. For the \emph{Planck} constraints we use 
a maximum $\ell$ of $\ell_{\rm max}=2500$.
We will present results for a fiducial weak lensing 
survey which is based on the \emph{DUNE} weak lensing concept, the 
assumed survey parameters are outlined in Table \ref{surveys}.
\begin{table}
\begin{ruledtabular}
\begin{tabular}{lc}
Survey&DUNE (fiducial)\\
Area/sqdeg&$20,000$\\
$z_{\rm median}$&$0.90$\\
$n_0$/sqarcmin&$35$\\
$\sigma_z(z)/(1+z)$&$0.025$\\
$\sigma_{\epsilon}$&$0.25$\\ 
\end{tabular}
\caption{The parameters describing the weak lensing survey investigated.}
\label{surveys}
\end{ruledtabular}
\end{table}
The $z_{\rm median}$ is the median redshift of the number density distribution of galaxies with redshift, and 
$n_0$ is the observed surface number density of galaxies. $\sigma_z(z)$ describes how the average 
accuracy with which a galaxies position in redshift is known, and $\sigma_{\epsilon}$ is the statistical 
variance of the intrinsic observed distortion of galaxies due to their random orientation.  
Note that we expect the photometric redshift error to have a small effect on the 
predicted statistical constraints as shown in \cite{art:Heavens06}, however a good photometric redshift 
error is required to reduce the 
effect of intrinsic alignment systematics as shown in \cite{art:Bridle} and \cite{art:Kitching08}. 

\subsection{The Fisher Matrix \& Bayesian Evidence}
\label{B Evidence}
The results presented in this paper will use the Fisher matrix formalism to make predictions of the cosmological 
parameter errors. The Fisher matrix is defined as the second derivative of the likelihood surface about the maximum.
In the case of Gaussian distributed data with zero mean this is given by 
\cite{art:Tegmark, art:Jungman, art:Fisher}
\be 
\label{fish}
F_{\alpha\beta}=\frac{1}{2}{\rm
Tr}\left[C^{-1}C_{,\alpha}C^{-1}C_{,\beta}\right]
\ee
where $C=S+N$ is the theoretical covariance of a particular method which consists of signal $S$ and noise $N$ 
terms. The commas in Eqn. (\ref{fish}) denote derivatives with respect to cosmological parameters about a 
fiducial cosmology. The fiducial cosmology used in this paper is based on the WMAP results 
\cite{art:Spergel};  
$\Omega_m=0.27$, $\Omega_{de}=0.73$, 
$\Omega_b=0.04$, $h=0.71$, $\sigma_8=0.80$, $w_0=-1.0$, $w_a=0.0$, $n_s=1.0$, $\alpha_n=0.0$. We consider two sets of fiducial value for the neutrino mass; one which is in agreement with 
current cosmological constraints, and one in which there are no massive neutrinos. This will 
allow for some discussion on the sensitivity of the predicted results to the fiducial values.
The first $m_{\nu}=0.66$ eV and $N_{\nu}=3.0$ 
is high compared to the expected hierarchical mass scale.
We justify this since we are using the current constraint on the neutrino mass from 
cosmology. The second set of fiducial values are $m_{\nu}=0$ eV 
and $N_{\nu}({\rm massless})=3.0$. 

The predicted marginal errors are calculated by taking the inverse of the 
Fisher matrix, the error on the $i^{\rm th}$ cosmological parameter is given by $\Delta\theta_i=\sqrt{(F^{-1})_{ii}}$. To combine constraints from multiple experiments the Fisher matrices are summed 
e.g $F_{\rm total}=F_{\rm lensing}+F_{\rm CMB}$.

In addition to presenting the marginal error on the cosmological parameters we will 
present the expected  
Bayesian \emph{evidence} that the fiducial survey could achieve for 
massive neutrinos. Computing the evidence allows one to \emph{distinguish} 
different models rather than 
constrain parameters \emph{within} a model (e.g. \cite{art:Hobson}) as explained below. 
A procedure for calculating the 
expected evidence directly from the Fisher matrix was presented in \cite{art:Heavens07}. 
In the case of massive neutrinos there is a 
natural question that an evidence calculation can answer: does the data provide evidence 
for massive neutrinos? 
Note that this is distinct from assuming that massive neutrinos exist 
and using the data to constrain their expected properties. 

The concept of evidence is derived from Bayes' theorem which relates the probability 
of model given the data $p(M|D)$, to the probability of the data given the model $p(D|M)$  
\be
p(M|D) = \frac{p(D|M)p(M)}{p(D)}
\end{equation}
where $p(M)$ is the prior probability on any parameters within the model $M$. We assume two competing models $M$ and 
$M'$. We also assume that $M'$ is a simpler model than $M$, 
containing fewer parameters $n'<n$, and that the models
are nested i.e. the more complex model $M$ is an extension of the simpler model $M'$. 
By marginalisation $p(D|M)$, known as the \emph{evidence}, is
\begin{equation}
p(D|M) = \int d\theta\,p(D|\theta,M)p(\theta|M).
\end{equation}
The posterior relative probabilities of the two models,
regardless of what their model parameters are, is
\begin{equation}
\frac{p(M'|D)}{p(M|D)}=\frac{p(M')}{p(M)}\frac{\int
d\theta'\,p(D|\theta',M')p(\theta'|M')}{\int
d\theta\,p(D|\theta,M)p(\theta|M)}.
\end{equation}
By assuming uniform priors on the models, $p(M')=p(M)$, this ratio
simplifies to the ratio of evidences which is called the {\em Bayes Factor},
\begin{equation}
B \equiv \frac{\int d\theta'\,p(D|\theta',M')p(\theta'|M')}{\int
d\theta\,p(D|\theta,M)p(\theta|M)}.
\end{equation}
It is the evaluation of the Bayes factor that allows one to determine whether a data set can distinguish between 
competing models. We wish to forecast whether a future survey will be able to distinguish between models. This can 
be done using the Fisher matrix using the following expression, given in \cite{art:Heavens07}
\begin{equation}
\label{BF}
B =
(2\pi)^{-p/2}\frac{\sqrt{\det{F}}}{\sqrt{\det{F'}}}\exp\left(-\frac{1}{2}\delta\theta_\alpha
F_{\alpha\beta}\delta\theta_\beta\right)\prod_{q=1}^p\Delta\theta_{n'+q},
\end{equation}
with $\delta\theta_\alpha$ given by 
\begin{equation}
\delta\theta'_\alpha =
-(F'^{-1})_{\alpha\beta}G_{\beta\zeta}\delta\psi_\zeta \qquad
\alpha,\beta=1\ldots n, \zeta=1\ldots p
\end{equation}
where $p\equiv n-n'$. Note that $F$ and $F^{-1}$ are $n \times n$ matrices, $F'$ is $n' \times n'$, and
$G$ is an $n' \times p$ block of the full $n \times n$ Fisher matrix $F$. 
$\delta\psi$ are the differences in the parameters values between models 
$M$ and $M'$. $\Delta\theta$ are any prior ranges imposed on the parameters we set 
$\Delta\theta=1$ at all times. 
The expression in Eqn. (\ref{BF}), given its implicit 
assumption of Gaussian likelihood surfaces, allows one to very quickly 
evaluate the expected evidence. This was 
done in \cite{art:Heavens07} for the case of modified gravity, 
forecasting the expected evidence 
for a single extra parameter $\gamma$ which parameterises any deviation 
from General Relativity. Here we 
will use Eqn. (\ref{BF}) to calculate the \emph{joint} 
evidence for two parameters $m_{\nu}$ and $N_{\nu}$. The Bayesian evidence has 
been extensively studied in 
cosmology in general 
(for example \cite{art:Trotta, art:Trotta07, art:Trotta07b, art:Marshall}) and in 
the field of dark energy (for example \cite{art:Liddle, art:Saini, art:Serra, art:Mukher}).

Throughout this paper we will use the Jeffreys \cite{art:Jeffreys} 
scale in which, $\ln B<1$ is 
`inconclusive', $1<\ln B<2.5$ is described
as `substantial' evidence in favour of a model, $2.5<\ln B< 5$ is `strong', 
and $\ln B>5$ is `decisive'.

\section{Marginal Error Results}
\label{Results}
In this Section we will present the predicted marginal errors on the massive neutrino 
parameters could be found using 3D cosmic shear in combination with a CMB \emph{Planck} 
experiment. 

Table \ref{marginalerrors} shows the marginalised cosmological parameter errors predicted 
for \emph{Planck} alone and combined with the 3D cosmic shear constraints from the fiducial 
survey. We also include the dark energy pivot redshift error. The pivot redshift 
is the point at which the error on $w_p$ minimises, this is defined by rewriting 
the dark energy parameterisation used $w(a)=w_0+w_a(1-a)$ to $w(a)=w_p+w_a(a_p-a)$ where 
$a_p=a_p(z_p)$ can correspond to any redshift. The Dark Energy Task Force (DETF; 
\cite{art:Albrecht}) Figure of Merit (FoM) is defined as being proportional  
to the reciprocal of the area constrained in the ($w_p$, $w_a$) plane at the pivot redshift 
\be
\label{FoM}
{\rm FoM}=1/\Delta w_p\Delta w_a. 
\ee
Note we use the reciprocal of the 
$1$-$\sigma$ two parameter ellipse, the DETF use the $2$-$\sigma$ two-parameter ellipse which differs from equation (\ref{FoM}) by a constant factor. 

By combining 3D cosmic shear constraints with \emph{Planck} the massive neutrino parameters 
could be constrained with marginal errors of $\Delta m_{\nu}\sim 0.03$ eV and 
$\Delta N_{\nu}\sim 0.08$ if the neutrinos are massive, 
this is a factor of $4$ improvement over \emph{Planck} alone. 
The dramatic improvement comes from the lifting of parameter degeneracies when the 
extra constraints are added. As shown in \cite{art:Kitching08} without massive neutrinos the 
fiducial survey design could provide a dark energy FoM$=475$. The inclusion of the massive 
neutrino parameters does not degrade this FoM substantially, 
since the extra parameters are well constrained. If there are no massive neutrinos
then the marginal errors on these parameters degrade, in this case the mass and number 
could be constrained to $\Delta m_{\nu}\sim 0.07$ eV and 
$\Delta N_{\nu}\sim 0.10$, however this is still a factor of $4$ improvement 
over \emph{Planck} alone. 

This degradation in the marginal error occurs because the effect 
of massive neutrinos on the matter power spectrum and hence on 3D weak lensing 
is non-linear. If the mass of neutrinos  
is larger then the amount of suppression at a given scale increases, furthermore the 
effect on the linear growth factor as a function of $f_{\nu}$ given by 
\cite{art:Kiakotou} is non-linear. In addition  
the scale at which power is suppressed due to free-streaming 
varies as a function of neutrino mass; neutrinos 
with lighter masses suppress growth at larger scales (higher-k) at a given redshift. 
\cite{art:Takada} investigated the effect of the neutrino mass fiducial model when making 
predictions on future marginal errors for a galaxy redshift survey and found that if 
the fiducial value of $f_{\nu}\ls 0.01$ then the marginal error on $N_{\nu}$ depends 
strongly on the assumed value of $f_{\nu}$ (e.g. \cite{art:Takada}, Figure 3).
One should only expect parameters fiducial values to have a small effect on the predicted 
marginal errors if they have a linear 
affect the covariance of a method i.e. so that the derivatives in the Fisher matrix 
(Eqn. \ref{fish}) do not change as the point around which the derivative is taken changes.
The assumed 
fiducial value of the neutrino parameters has a small effect 
on the errors of most other cosmological parameters, given that the fiducial models are 
so different, except for $\Omega_m$. This is because 
the contribution of massive neutrinos to the total mass-energy behaves with redshift like a 
extra matter density. 

Hannestad et al. (2006) \cite{art:Hannestad06} find that, 
by combining the weak lensing tomography 
constraints from their Wide-5 survey (which is similar to our fiducial survey) 
with \emph{Planck}, $\Delta m_{\nu}=0.043$ eV and $\Delta N_{\nu}=0.067$. They assume 
a massive neutrino fiducial model with $m_{\nu}=0.07$ eV and $N_{\nu}=3.0$. 
We find good agreement between this result and our massive neutrino fiducial model 
despite using different weak lensing methods, slightly different survey 
designs and different fiducial models. 
They find a \emph{Planck}-only predicted errors of $\Delta m_{\nu}=0.48$ eV 
and $\Delta N_{\nu}=0.19$ although they do not include B-modes which we do in 
our CMB Fisher matrix \cite{art:Taylor}, we also 
checked the CMB Fisher matrices with an MCMC analysis and found that the error 
on $N_{\nu}$ for \emph{Planck}, assuming 
that the neutrino mass is fixed, was $\Delta N_{\nu}=0.084$ and for the Fisher 
matrix analysis we find $\Delta N_{\nu}=0.10$ assuming that the mass is fixed. 
\begin{table*}
\begin{ruledtabular}
\begin{tabular}{lcc|cc}
Fiducial values:&$m_{\nu}=0.66$ eV \& $N_{\nu}=3.00$&&$m_{\nu}=0$ eV \& $N_{\nu}=3.00$&\\
\hline
Parameter&\emph{Planck} Alone&3D Cosmic Shear + \emph{Planck}&\emph{Planck} Alone&3D Cosmic Shear + \emph{Planck}\\
$\Delta\Omega_m$   &$ 0.0014 $&$0.0008$&$0.0104$&$0.0041$\\
$\Delta\Omega_{de}$   &$0.0015$  &$0.0012$&$0.0041$&$0.0021$\\
$\Delta h$         &$ 0.0167 $&$0.0055$&$0.0148$&$0.0060$\\
$\Delta\sigma_8$   &$ 0.0965 $&$0.0040$&$0.0999$&$0.0202$\\
$\Delta\Omega_b$   &$ 0.0019 $&$0.0006$&$0.0014$&$0.0007$\\
$\Delta w_0$         &$ 0.5622 $&$0.0442$&$0.6031$&$0.0309$\\
   $\Delta w_a$        &$ 1.8679 $&$0.2277$&$1.9158$&$0.1853$\\
  $\Delta n_s$         &$ 0.0103 $&$0.0018$&$0.01435$&$0.0039$\\
   $\Delta\alpha_n$   &$0.0083 $&$0.0044$&$0.0074$&$0.0046$\\
   $\Delta r$       &$ 0.0199 $&$0.0193$&$0.0207$&$0.0202$\\
   $\Delta\tau$       &$ 0.0084$ &$0.0078$&$0.0080$&$0.0078$\\
   $\Delta m_{\nu}$/eV   &$0.2815$&$0.0324$&$0.3815$&$0.0728$\\
   $\Delta N_{\nu}$   &$0.1144$ &$0.0836$&$0.2807$&$0.1042$\\
\hline
$\Delta w_p$&$0.1177$&$0.011$&$0.1879$&$0.011$\\
DETF FoM&$5$&$400$&$3$&$490$\\
\end{tabular}
\caption{Predicted marginalised 
  cosmological parameter errors for \emph{Planck} alone and combined with the 
  3D cosmic shear constraints from the fiducial survey. We also show the dark energy pivot 
  redshift error and the DETF Figure of Merit, FoM. We show results assuming two different 
  sets of fiducial values for the massive neutrino parameters one in which neutrino have mass
  and one in which neutrinos are massless.}
\label{marginalerrors}
\end{ruledtabular}
\end{table*}

As shown in  \cite{art:Kitching08} it can be expected that realistically 
the dark energy FoM constraints from 3D cosmic shear combined with a 
\emph{Planck} prior will be reduced by 
approximately a factor of $2$ due to photometric, intrinsic alignment and shape measurement 
systematic effects, this correspond to a factor of $\sqrt{2}$ 
for each dark energy parameter. Since, as highlighted by \cite{art:Kiakotou}, 
the neutrino mass parameters 
affect the power spectrum in a similar way to dark energy one should realistically 
expect \emph{at most} a factor of $\sqrt{2}$ reduction in the combined constraints 
due to systematics. 
Using this heuristic approximation this yields constraints 
of $\Delta m_{\nu}\sim 0.04$ eV and $\Delta N_{\nu}\sim 0.11$ for a massive neutrino fiducial
model. 

\section{Bayesian Evidence Results}
\label{Bayesian Evidence Results}
In the interpretation of the results in the following Section 
one should keep in mind that 
the magnitude of the Bayes factor shown 
is a prediction of an experiments \emph{ability to distinguish} 
one model over another i.e. to what 
level could the experiment in consideration will be able to provide 
evidence for the fiducial model over 
another competing model (or vice versa) where the models are distinguished by changes in the 
parameter values $\delta\psi_i$.

Note that in the evidence calculation we use the $m_{\nu}=0$ eV 
and $N_{\nu}({\rm massless})=3.0$ fiducial values for the Fisher matrices since 
this represents the `simple' model as described in Section \ref{B Evidence} and will allow 
for statements to be made about whether the data could provide evidence for a more 
complicated model containing massive neutrinos over a simpler model 
with no massive neutrinos.

\subsection{Multi-Parameter Expected Evidence}
\begin{figure}
\resizebox{0.8\columnwidth}{!}{
\includegraphics{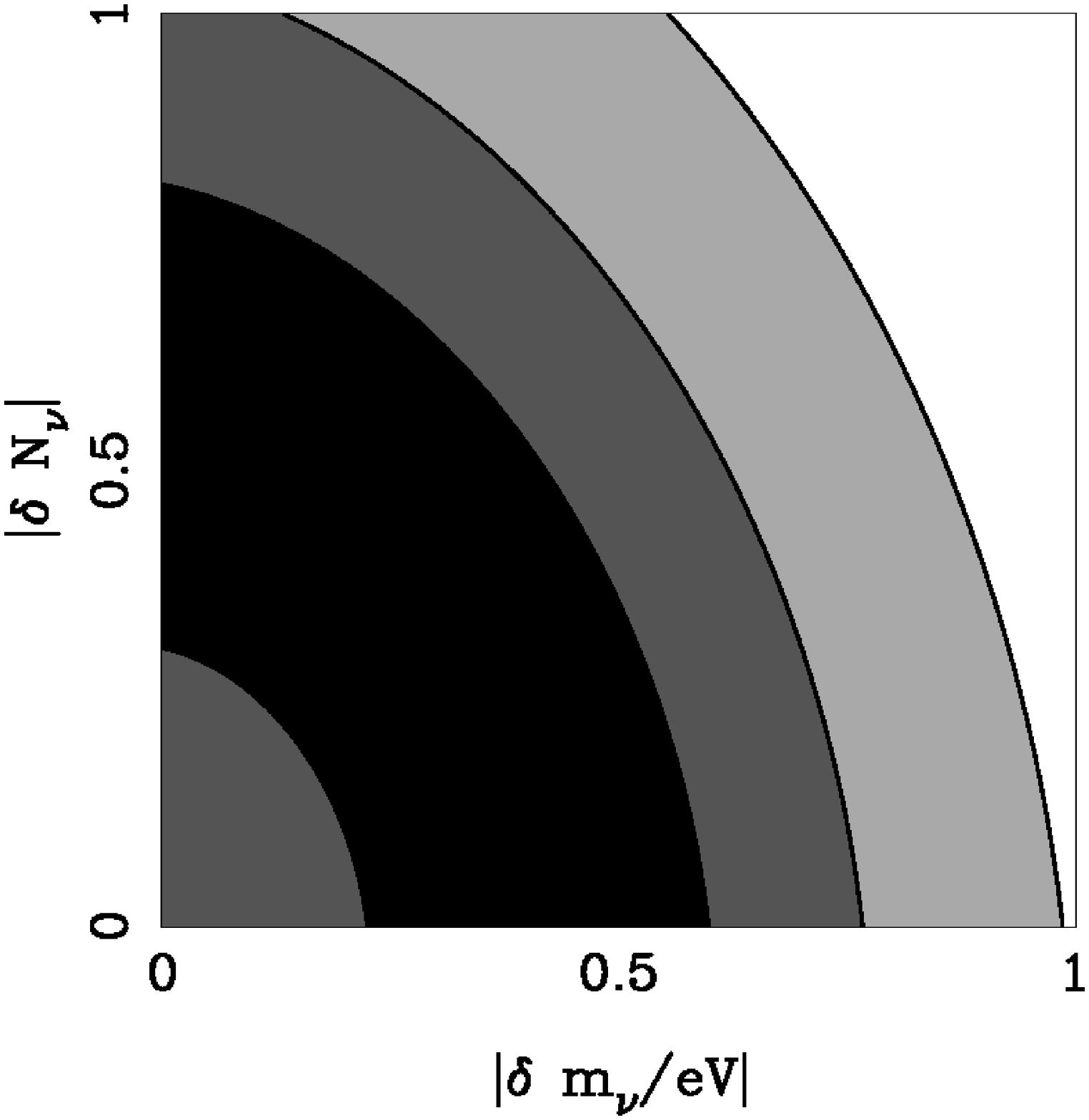}}

\vspace{2mm}

\resizebox{0.8\columnwidth}{!}{
\includegraphics{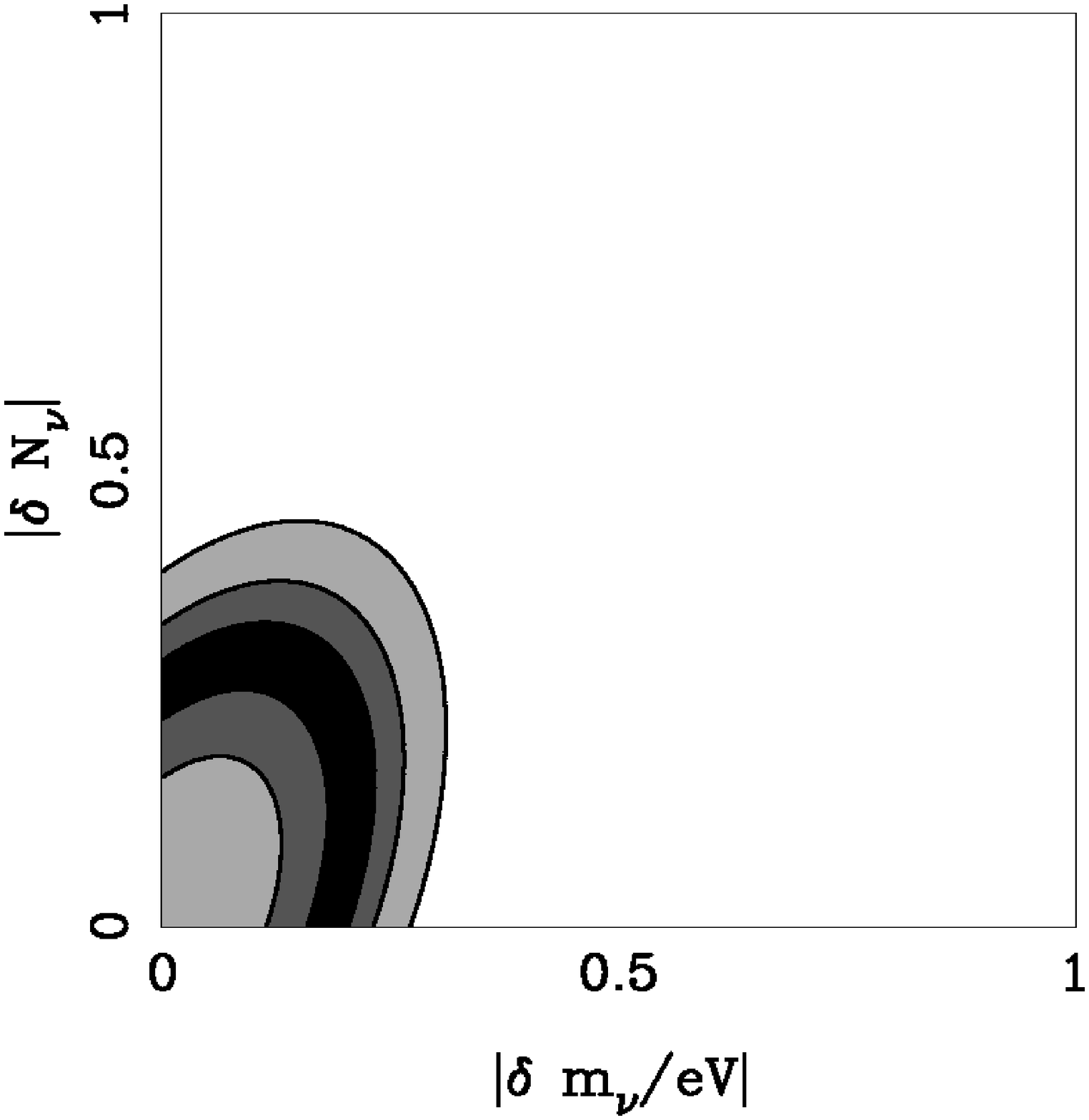}}

\caption{The expected joint evidence for the number 
$N_{\nu}$ and mass $m_{\nu}$ of neutrinos
using 3D cosmic shear and the fiducial survey design. 
White=`decisive', lightest gray=`substantial', darkest gray=`strong' 
and black=`inconclusive'. The upper panel shows the constraints 
from \emph{Planck} alone, the lower panel shows the constraints when 3D cosmic shear from the 
fiducial survey is combined with \emph{Planck}.}
\label{figure2D}
\end{figure}

Fig. \ref{figure2D} shows the expected 
evidence contours for $m_{\nu}$ and $N_{\nu}$ jointly for the 
fiducial survey design. Note that a $\delta N_{\nu}=0$ means that $N_{\nu}=3$ 
and $\delta m_{\nu}=0$ means that $m_{\nu}=0$ eV i.e. at the fiducial values. 
The figure shows that there is a substantial improvement in combing 
\emph{Planck} with 3D weak lensing data. On its own Planck could only 
provide at most  
substantial evidence massive neutrinos  
for models with a large range of massive neutrinos parameter values. 
The fiducial survey using 3D cosmic shear 
combined with a \emph{Planck} prior will; 
\begin{itemize}
\item
Provide substantial evidence for massive neutrinos over models in which there are no 
massive neutrinos, and if the neutrino mass is small $\delta m_{\nu} \ls 0.1$ eV then there 
will be substantial evidence for these models. 
\item 
Be able to decisively distinguish between models in which there are no massive 
neutrinos and models in which $N_{\nu}\ls 3.00-0.40$ or $N_{\nu}\gs 3.00+0.40$ and 
$m_{\nu} \gs 0.25$ eV. 
\item
Specifically the experiment could decisively 
distinguish between models in which there are $3$ massless neutrinos and 
i) models in which there are few $N_{\nu}<2.6$ 
(possibly zero) massive $m_{\nu}>0.25$ eV neutrinos 
ii) models in which there are many $N_{\nu}>3.4$ massive $m_{\nu}>0.25$ eV neutrinos 
iii) models in 
which there are few $N_{\nu}<2.6$ massless $m_{\nu}=0$ eV neutrinos and 
iv) models in which there are many $N_{\nu}>3.4$ massless  $m_{\nu}=0$ eV neutrinos.
\end{itemize}
There is a band in which the 
evidence is inconclusive (the black band in Fig. \ref{figure2D}), this represents the 
boundary between where the data would favour the simpler fiducial model and the situation 
in which the data would 
favour a different model (i.e. where the probability of either the 
fiducial or a different model 
being correct is equal). 

\subsection{Single-Parameter Expected Evidence}
\begin{figure}
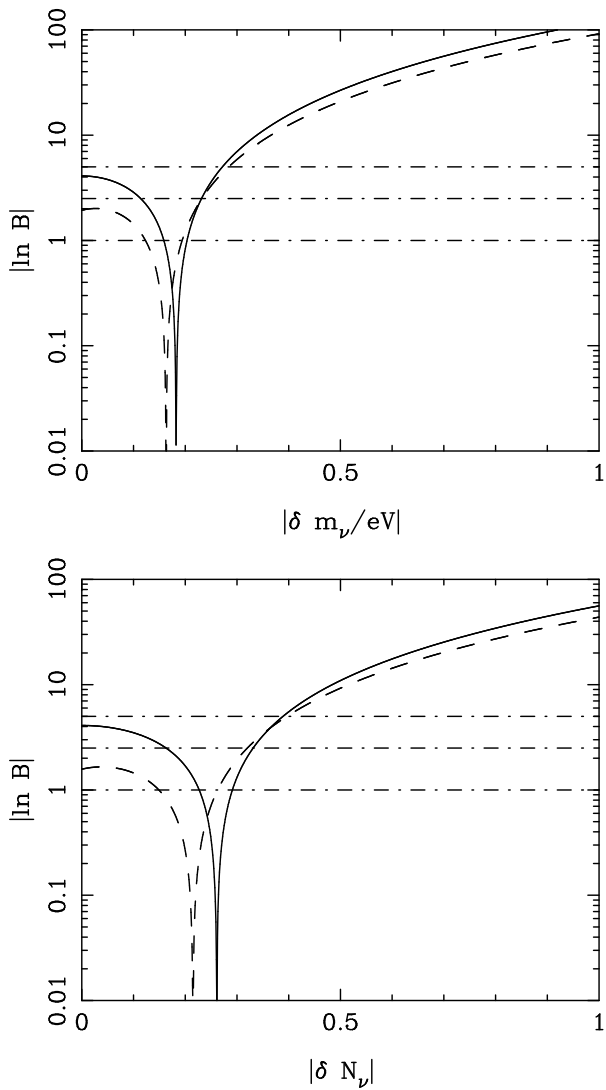

\resizebox{0.925\columnwidth}{!}{
\includegraphics{fig1c.ps}}

\vspace{2mm}

\resizebox{0.925\columnwidth}{!}{
\includegraphics{fig1d.ps}}
\caption{The predicted evidence for the number $N_{\nu}$ and total mass $m_{\nu}$ of 
  neutrinos
  individually for 3D cosmic shear using the fiducial survey combined with a 
  \emph{Planck} prior.
  In each plot the solid line show the conditional evidence assuming that the 
  other parameter is fixed at its fiducial value, the dashed line shows the 
  marginal expected 
  evidence when the possible values of the hidden parameter are taken into account, see 
  Eqn. (\ref{marginalevidence}). The dot-dashed lines show the defining evidence limits on 
  the Jeffery's scale where $\ln B<1$ is 
  `inconclusive', $1<\ln B<2.5$ is `substantial' evidence in favour of a model, 
  $2.5<\ln B< 5$ is `strong', and $\ln B>5$ is `decisive'.}
\label{figure1D}
\end{figure}

As well as the joint expected evidence on the two massive neutrino parameters we can also 
investigate the expected evidence for either parameter individually. When this is done there 
are two ways in which the other (hidden) parameter(s) can be dealt with; 
\begin{itemize}
\item
The hidden parameters can either be assumed to 
be fixed at their fiducial values. We will refer 
to this as the \emph{conditional} evidence. In this case the expected 
evidence presented has the
implicit assumption that the hidden parameters have the value chosen.
\item
The evidence can be integrated over the hidden parameters to obtain what we will 
refer to as 
the \emph{marginal} evidence. For a multi-parameter model which depends on 
$\theta_{i=1,...,n}$ 
parameters the total expected evidence is a function of all these 
parameters $B(\theta_{i=1,...,n})$. 
The marginal evidence on one of the parameter $\theta_j$ is given by 
\be 
\label{marginalevidence}
B(\theta_j)=\int d\theta_1...d\theta_{j-1}d\theta_{j+1}...d\theta_n B(\theta_{i=1,...,n}).
\ee
In the case presented here, that of two parameters, the total evidence $B(N_{\nu},m_{\nu})$ 
can be integrated to obtain the marginal evidence $B(N_{\nu})$ or $B(m_{\nu})$.
\end{itemize}
Fig. \ref{figure1D} shows the one-dimensional expected evidence for 3D cosmic shear combined 
with a \emph{Planck} prior. It can be seen from both panels in this Figure that by using the 
conditional evidence one can over estimate the evidence when the deviation between models 
is small by upto a factor of $\sim 5$, 
however when the models being compared are very different (large values of $\delta\psi$)
the marginal and conditional evidences converge. The results drawn from 
these plots are similar to
those from the full joint evidence. 3D cosmic shear should find substantial evidence that
neutrinos are massless if this is the case. Furthermore this experiment could decisively 
distinguish between models in which there are no massive neutrinos and models in 
which there are massive neutrinos with $m_{\nu}\gs 0.25$ eV, 
or if the number of neutrinos differs by $|N_\nu-3|\gs 0.35$   

\section{Conclusion}
\label{Conclusion}
In this paper we have shown that 3D cosmic shear has the ability to measure the 
effect that massive neutrinos 
can have on the matter power spectrum, and use this effect to place constraints on 
the total mass of neutrinos $m_{\nu}=\sum_i m_i$ and 
number of these massive neutrinos $N_{\nu}$. By combining the results using 3D cosmic shear 
from a next generation 
photometric redshift survey with the constraints from the \emph{Planck} CMB experiment 
one could expect marginalised errors for the massive neutrino parameters of 
$\Delta m_{\nu}\sim 0.03$ eV and $\Delta N_{\nu}\sim 0.08$ which is a factor of 
$4$ improvement over the 
constraints using the CMB alone. We found that if one assumes in this calculation 
that neutrinos are massless then the predicted marginal error on these parameters is 
substantially degraded to 
$\Delta m_{\nu}\sim 0.07$ eV and $\Delta N_{\nu}\sim 0.10$, however this is still an 
improvement of a factor of $4$ over the marginal errors from \emph{Planck} alone using 
the same fiducial model. This increase in the marginal errors occurs because the power 
spectrum is affected by neutrino mass in a non-linear way.

Even by including heuristically including systematic effects, using a rule-of-thumb 
from Kitching et al. (2008), the 
improvement over \emph{Planck} alone 
is still a factor of $3$. Comparing with other probes we find 
that 3D weak lensing is competitive; \cite{art:Takada} find that using a 
galaxy redshift survey combined with a \emph{Planck} prior $\Delta m_{\nu}\gs 0.025$ eV, 
\cite{art:Gratton} find that in combination with \emph{Planck} constraint 
Lyman-$\alpha$ experiments 
could constrain $\Delta m_{\nu}\ls 0.06$ eV. However we note that these constraints 
should be dependent on the fiducial value of the neutrino mass chosen. 

We explicitly presented results for a fiducial survey which has the 
characteristics of \emph{DUNE} however 
other forthcoming surveys are also well suited to do 3D weak lensing and 
should have a similar
sensitivity to neutrino mass for example Pan-STARRS-1 
should yield constraint roughly twice that of \emph{DUNE} \cite{art:Kitching08} 
$\Delta m_{\nu}\sim 0.06$ eV and $\Delta N_{\nu}\sim 0.16$. The LSST should 
yield constraints of roughly the same 
order of magnitude as \emph{DUNE}.

Using the expected evidence calculation from \cite{art:Heavens07} 
we have shown that one can expect 
substantial evidence for massive neutrinos if they exist, and 
furthermore that one could decisively distinguish
between models in which there are no massive neutrinos and models in which there are 
massive neutrinos with  
$|N_\nu-3|\gs 0.35$ and $m_{\nu}\gs 0.25$ eV.  

We have introduced the concept of 
marginal and conditional evidence and shown that by assuming the value of one parameter in a model 
to be fixed the one-parameter evidence can be under or over estimated by upto a factor $5$. These evidence calculations can 
be generalised to models with an arbitrary number of parameters, and the simple application of this  
algorithm using only the Fisher matrix can allow predictions to be made which would be prohibitively 
time consuming using traditional evidence calculations (with the caveat that Gaussianity is assumed).

If the constraints predicted in this paper were realised then our understanding of massive neutrinos could be 
revolutionised allowing the physics beyond the Standard Model which this implies to be understood more 
entirely. 

\section*{Acknowledgments}
TDK is supported by the Science and Technology Facilities Council,
research grant number E001114. LV is supported
by FP7-PEOPLE-2007-4-3-IRG n202182, and by  CSIC I3 grant :
200750I034. We thank Markus Ahlers and 
the DUNE weak lensing working group for useful discussions.


\begin{thebibliography}{}

\bibitem{art:Abazajian} Abazajian K.; Dodelson S.; 2003, Phys.Rev.Lett., 91, 041301 
\bibitem{art:Abdalla} Abdalla, F.; Rawlings S.; 2007, MNRAS, 381, 1313
\bibitem{art:Ahmed} Ahmed S.; et al. [SNO Collaboration]; 2003, arXiv:nucl-ex/0309004.
\bibitem{art:Ahn} Ahn M.; et al. [K2K Collaboration]; 2006, Phys. Rev. D, 74, 072003
\bibitem{art:Bacon} Bacon, D.; et al.; 2003; MNRAS, 363, 723-733
\bibitem{art:Bridle} Bridle, S.; King, L.; eprint arXiv:0705.0166
\bibitem{art:Albrecht} Albrecht, A.; et al.; eprint arXiv:astro-ph/0609591
\bibitem{art:Castro} Castro, P. G.; Heavens, A. F.; Kitching, T. D.; 2005, PhRvD, 72, 3516
\bibitem{art:Chevallier} Chevallier, M.; Polarski, D.; 2001, IJMPD, 10, 213
\bibitem{art:Cooray} Cooray A.; 1999, Astron.Astrophys., 348, 31
\bibitem{art:Eguchi} Eguchi K.; et al. [KamLAND Collaboration]; 2003, Phys. Rev. Lett., 90, 021802
\bibitem{art:Eisenstein} Eisenstein D.; Hu W.; 1998, ApJ, 496, 605
\bibitem{art:Fisher} Fisher, R.; 1935, JRoyStatSoc, 98, 35 
\bibitem{art:Free} Friedland, A.; Zurek, K. M.; Bashinsky, S.; 2007, astro-ph, arXiv:0704.3271
\bibitem{fogli04} Fogli G., Lisi E., Marrone A., Melchiorri A., Palazzo A., Serra P., Silk J., Phys.\ Rev.\ D, 70 (2004) 113003 [arXiv:hep-ph/0408045]
\bibitem{art:Fogli:2006yq} Fogli G., et al., 2007, Phys.Rev.D, 75, 053001, [arXiv:hep-ph/0608060]
\bibitem{art:Fukuda} Fukuda Y.; et al. [Super-Kamiokande Collaboration]; 1998, Phys. Rev. Lett., 81, 1562 
\bibitem{art:Fukugita} Fukugita M.; Yanagida T.; 1986, Phys. Lett. B, 174, 45
\bibitem{art:Gratton} Gratton S.; Lewis A.; Efstathiou G.; 2007, arXiv:0705.3100v1 [astro-ph]
\bibitem{art:Hannestad07} Hannestad S.; Wong Y.; 2007, JCAP, 0707, 004
\bibitem{art:Hannestad06} Hannestad S.; Tu H.; Wong Y.; 2006, JCAP 0606, 025
\bibitem{art:Heavens00} Heavens, A. F.; Refregier, A.; Heymans, C.; 2000, MNRAS, 319, 649
\bibitem{art:Heavens03} Heavens, A. F.; 2003, MNRAS, 343, 1327
\bibitem{art:Heavens06} Heavens, A. F.; Kitching, T. D.; Taylor, A. N.; 2006, MNRAS, 373, 105
\bibitem{art:Heavens07} Heavens, A. F.; Kitching, T. D.; Verde L.; 2007, MNRAS, 380, 1029
\bibitem{art:Hobson} Hobson, M.; Bridle S.; Lahav, O.; 2002, MNRAS, 335, 377
\bibitem{art:Itow} Itow Y.; et al. [The T2K Collaboration]; 2001, arXiv:hep-ex/0106019.
\bibitem{art:Jeffreys} Jeffreys H.; 1961, Theory of Probability, Oxford University Press
\bibitem{art:Jungman} Jungman, G.; Kamionkowski, M.; Kosowsky, A.; Spergel, D.; PRD, 54, 1332
\bibitem{art:Kaiser} Kaiser N., et al; 2002, SPIE, 4836, 154
\bibitem{art:Kiakotou} Kiakotou A.; Elgaroy O.; Lahav O.; 2007, arXiv:0709.0253 [astro-ph]
\bibitem{art:King} King, S., 2007; arXiv:0712.1750v1 [physics.pop-ph]
\bibitem{art:Kitching07} Kitching, T. D.; Heavens, A. F.; Taylor, A. N.; Brown, M. L.; Meisenheimer, K.; Wolf, C.; Gray, M. E.; Bacon, D. J.; 2007, MNRAS, 376, 771
\bibitem{art:Kitching08} Kitching, T. D.; Taylor, A. N.; Heavens, A. F.; 2008a, submitted to MNRAS
\bibitem{art:Kristiansen} Kristiansen J.; Elgaroy O.; 2007, arXiv:0709.4152 [astro-ph]
\bibitem{art:Liddle} Liddle A.; Mukherjee P.; Parkinson D.; Wang Y.; 2006, Phys.Rev. D, 74, 123506
\bibitem{art:Lesgourgues} Lesgourgues J.; Pastor S.; Phys.Rept., 429, 307-379
\bibitem{Troi}  V.M.~Lobashev, in the Proceedings of NPDC 17,
                ed.\ by N.\ Auerbach, Zs.\ Fulop, Gy.\ Gyurky, and E.\
                Somorjai, Nucl.\ Phys.\ A 719, 153 (2003).
\bibitem{art:Loveday} Loveday J.; et al. [the SDSS Collaboration]; 2002, Contemp. Phys., 43, 437
\bibitem{art:LSND} The LSND Collaboration (A. Aguilar et al.), Phys. Rev. D64, 112007 (2001).
\bibitem{Ma95}  C.P.~Ma and E.~Bertschinger Astrophys.\ J.\ 455, 7 (1995).
\bibitem{art:Marshall} Marshall P.; Rajguru N.; Slosar A.; 2006, Phys.Rev. D, 73, 067302
\bibitem{art:Massey} Massey R.; et al.; 2007, ApJS, 172, 239
\bibitem{Mc04}  SDSS Collaboration, P.~McDonald {\it et al.},astro-ph/0405013.
\bibitem{Mbet}  B.H.J.\ McKellar,
                Phys.\ Lett.\ B {97}, 93 (1980);
                F.~Vissani, Nucl.\ Phys.\ B (Proc.\ Suppl.)
		{100}, 273 (2001);
                J.\ Studnik and M.\ Zralek, hep-ph/0110232. See also
                the discussion in Y.~Farzan and A.Yu.~Smirnov,
                Phys.\ Lett.\ B {557}, 224 (2003).
\bibitem{art:MB} The MiniBooNE Collaboration (E. Church et al.) FERMILAB-P-0898 (1997),available at http://library.fnal.gov/archive/test-proposal/0000/fermilab-proposal-0898.shtml.
\bibitem{art:Mukher} Mukherjee, P.; Parkinson, D.; Corasaniti, P. S.; Liddle, A.; Kunz; 2006,  MNRAS, 369, 1725 
\bibitem{art:Munshi} Munshi, D.; Valageas, P.; Van Waerbeke, L.; Heavens, A.; eprint arXiv:astro-ph/0612667
\bibitem{art:Osipowski} Osipowicz A.; et al. [KATRIN Collaboration]; 2001, arXiv:hep-ex/0109033.
\bibitem{art:Peacock} Peacock, J.; Schneider, P.; 2006, Msngr, 125, 48
\bibitem{art:Peiris} Peiris H.; 2005, Contemp. Phys., 46, 77
\bibitem{art:Refregier} Refregier A.; et al.; 2006, SPIE, 6265, 58 
\bibitem{art:Saini} Saini T.D; Weller J.; Bridle S.; 2004, MNRAS, 348, 603
\bibitem{art:Saito} Saito, S.; Takada, M.; Taruya, A.; 2008, astro-ph, arXiv:0801.0607
\bibitem{art:Seljak96} Seljak U.; Zaldarriaga M.; 1996, ApJ, 469, 437
\bibitem{art:Seljak05} Seljak U.; et al.; 2005, Phys.Rev. D, 71, 103515
\bibitem{art:Seljaks} Seljak, U.; Makarov, A.; McDonald, P.; Trac, H.; 2006, Phys.Rev.Lett., 97, 191303
\bibitem{art:Serra} Serra P.; Heavens A.; Melchiorri A.; 2007, MNRAS, 379, 1, 169
\bibitem{art:Smith} Smith R.; et al.; 2003, MNRAS, 341, 1311
\bibitem{art:Spergel} Spergel D., et al.; 2007, ApJS, 170, 377
\bibitem{art:Taylor} Taylor, A. N.; et al.; 2004, MNRAS, 353, 1176
\bibitem{art:Takada} Takada, M.; Komatsu E.; Futamase T.; 2006, Phys.Rev. D, 73, 083520
\bibitem{art:Tegmark} Tegmark, M.; Taylor A., Heavens A.; 1997, ApJ, 440, 22
\bibitem{art:Trotta} Trotta, R.; 2003, New Astron.Rev.,47, 769, 774
\bibitem{art:Trotta07} Trotta, R.; 2007, MNRAS, 378, 72-82
\bibitem{art:Trotta07b} Trotta, R.; 2007, MNRAS, 378, 819--824
\bibitem{art:PDG}  Yao W.-M.; et al.; 20003, J. Phys. G, 33, 1
\bibitem{Main}  C.\ Weinheimer, Nucl. Phys. Proc. Suppl. 118, (2003) 279.
\end{thebibliography}
\end{document}